\documentclass[prb,twocolumn,superscriptaddress,showpacs,amsmath,amssymb]{revtex4}
\usepackage{amsfonts}
\usepackage{bbm}
\usepackage{graphicx}
\usepackage{dcolumn}
\usepackage{bm}

\begin{document}

\title{Double refraction and spin splitter in a normal-hexagonal semiconductor junction}

\author{Peng Lv}
\affiliation{International Center for Quantum Materials, School of Physics, Peking University, Beijing 100871,
China}
\author{Ning Dai}
\affiliation{International Center for Quantum Materials, School of Physics, Peking University, Beijing 100871, China}
\author{Qing-Feng Sun}
\email[]{sunqf@pku.edu.cn}
\affiliation{International Center for Quantum Materials, School of Physics, Peking University, Beijing 100871, China}
\affiliation{Collaborative Innovation Center of Quantum Matter, Beijing, 100871, China}

\begin{abstract}
In analogy with light refraction at optical boundary, ballistic electrons also
undergo refraction when propagate across a semiconductor junction.
Establishing a negative refractive index in conventional optical materials is difficult,
but the realization of negative refraction in electronic system is conceptually straightforward,
which has been verified in graphene p-n junctions in recent experiments.
Here, we propose a model to realize double refraction and double focusing of electric current
by a normal-hexagonal semiconductor junction.
The double refraction can be either positive or negative, depending on the junction being n-n type or p-n type.
Based on the valley-dependent negative refraction, a spin splitter (valley splitter)
is designed at the p-n junction system,
where the spin-up and spin-down electrons are focused at different regions.
These findings may be useful for the engineering of double lenses in electronic system
and have underlying application of spin splitter in spintronics.
\end{abstract}
\maketitle

\section{Introduction}
The propagation of electrons has many similarities with the propagation of light.\cite{Datta,W.van}
In two-dimensional electron gas (2DEG), where the mean free path is larger
than the size of the system, ballistic electrons propagate following straight-line trajectories
which is analogous to light rays.
When the ballistic electrons transmit across a semiconductor junction, electrons should undergo
refraction in analogy with light refraction at optical boundary with different refractive indices.\cite{U.Sivan,H.van,J.Spector,Lv}
Such phenomena can be understood simply in terms of Snell's law, where the refractive index of photons is replaced by the wave vector
of electrons.
Thus, it is possible to manipulate electrons like photons and electron optics has attracted worldwide attention because of its underlying applications.
Electron focusing, diffraction, and double-slit interference experiments are examples of electron optics which have been clearly observed in 2DEG systems.\cite{add1,add2}

In conventional 2DEG systems, electrostatic lenses have been demonstrated in high mobility GaAs about thirty years ago.\cite{U.Sivan,J.Spector}
Since then, many work have been undertaken to obtain various electron optical devices like mirrors, prisms, lenses and splitters.\cite{A1,A2,A3,A4,A5}
One interesting topic of electron optics is the negative refraction,\cite{B1,B2}
which is challenging to achieve in conventional optical systems. For photons, this behavior can be realized in optical metamaterials.\cite{B3,add3}
In electronic systems, negative refraction can be achieved quite straightforward.\cite{A1,B4}
For example, when electrons transmit across a p-n junction, in order to conserve the transverse component of momentum, the transverse group velocity has to change a sign between the valence bands in p side and the conduction bands in n side, hence leads to the negative refraction.
Graphene has a unique band structure with a linear dispersion relation of low-lying
excitations, which gives rise to many peculiar properties.\cite{c3,ref1} Because of the vanishing band gap and the high intrinsic mobility, graphene has also been considered as an attractive platform for studying the electron optics.\cite{A1,A2,A3,A4,c1,c2,c4,ref2}
Recent experiments have clearly demonstrated the negative refraction in graphene p-n junctions.\cite{B4,c4}
This negative refraction can be used to design a perfect lens, and has many other potential applications.

In optics, a beam of light at the anisotropic crystal interface exhibits the double
refraction effects. Recently, monolayer transition-metal dichalcogenides (TMDs) have been successfully fabricated in experiments.\cite{IsSC1,IsSC2,IsSC3,IsSC4,IsSCth}
The TMDs (e.g. NbSe$_2$, MoS$_2$) exhibit Ising pairing in superconducting phase at sufficiently low temperature, with an in-plane
upper critical field far above the Pauli paramagnetic limit.\cite{IsSC1,IsSC2,IsSC3,IsSC4} This unusual behavior is attributed to the inter-valley pairing protected
by Zeeman-type spin-valley locking against external magnetic fields.\cite{IsSC2} This out-of-plane Zeeman-type spin polarization of the valleys can
be used to achieve double refraction in electron optics. Motivated by this,
in this paper, we propose a model to realize double refraction and double focusing
of electric current by a normal-hexagonal semiconductor junction.
This model is based on a hexagonal lattice system (like graphene or TMDs), but breaks the A-B lattice symmetry.
By introducing a Rashba spin-orbit interaction into the system,
the two valleys in the Brillouin zone are no longer equivalent.
When an electron is incident from the normal conductor side with the square lattice,
double refraction and double focusing occur, and the electron is transmitted to two different modes.
The two modes have different refractive indices due to the two valleys being inequivalent.
We investigate electron optics in both n-n junction and p-n junction below.
In particular, the incident electron undergoes double negative refraction in p-n junction.
We also show that by double negative refraction with elaborately tuning the chemical potential of the system,
the p-n junction can work as a spin splitter (valley splitter), which may be useful in spintronics
(valleytronics).

The rest of this paper is organized as follows.
In Sec. \ref{sec:Hamiltonian}, we present the model Hamiltonian and the corresponding band structures of the normal-hexagonal semiconductor junction, considering both the effect of
A-B lattice symmetry breaking and Rashba spin-orbit interaction.
In Sec. \ref{sec:Double refraction}, we investigate the positive double refraction in n-n junction and the negative double refraction in p-n junction.
In Sec. \ref{sec:Spin splitter}, a spin splitter is designed based on the negative double refraction.
Finally, a brief summary is presented in Sec \ref{sec:conclusions}.

\begin{figure}[!htb]
\includegraphics[width=1.0\columnwidth]{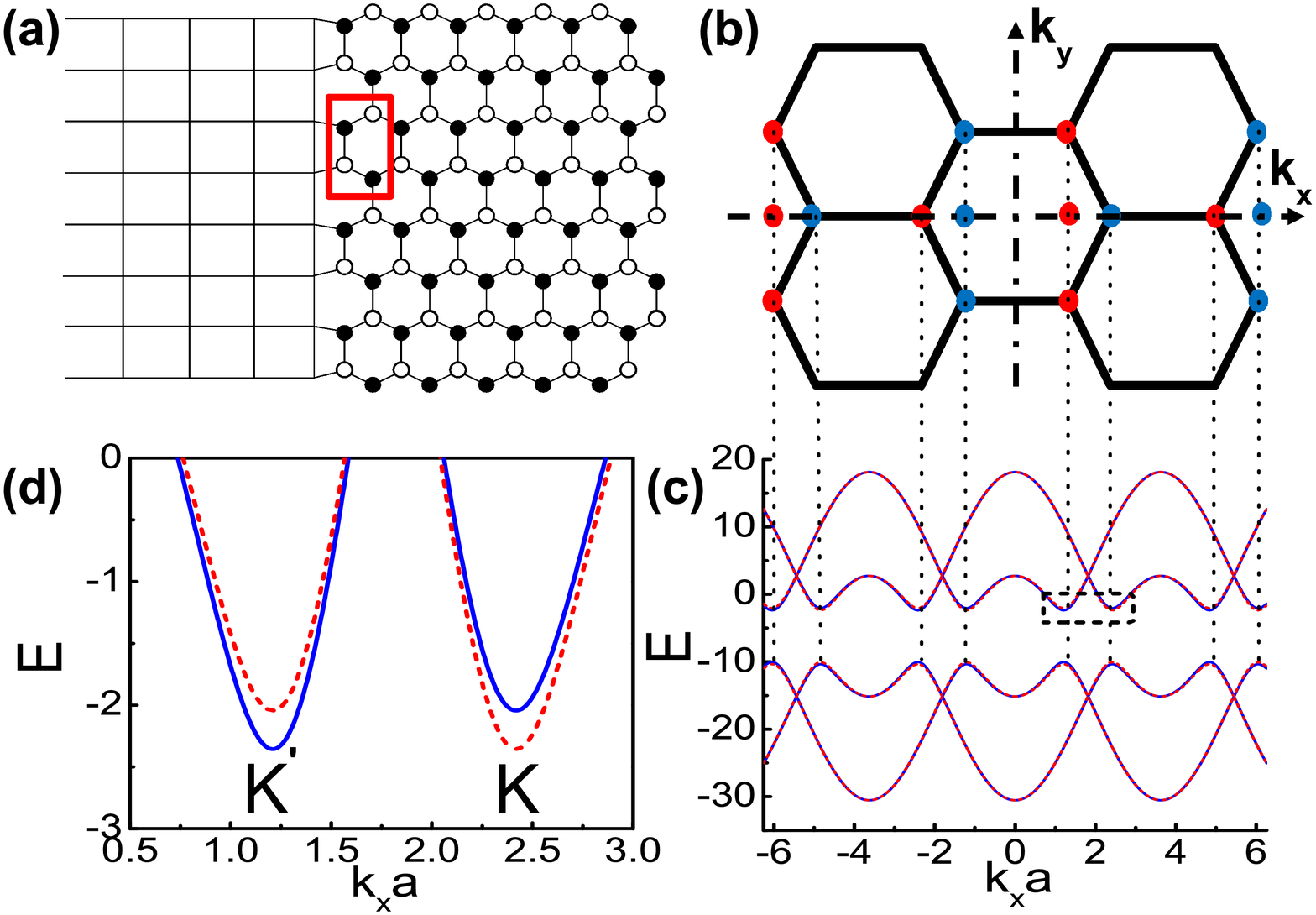}
\caption{(Color online) Lattice diagram and band structures.
(\textbf{a}) The schematic diagram of the normal-hexagonal semiconductor junction in the lattice model.
(\textbf{b}) Brillouin zone of the hexagonal lattice.
(\textbf{c}) Energy bands along the dash line in (\textbf{b}) in which $k_y=0$.
Dashed red lines denote the energy bands of spin-up electrons, while the blue lines are the spin-down bands.
(\textbf{d}) Zoom-in figure of the energy bands of the dotted box in (\textbf{c}).
The parameters are $\epsilon_{R}=-6.2$, $t_{R} =-8$, $\beta=4$, and $\beta_{s}=-0.03$.}
\label{fig1}
\end{figure}

\section{\label{sec:Hamiltonian}Model Hamiltonian and band structures}
In the following, we introduce the model Hamiltonian and demonstrate the corresponding band structures.
Fig. \ref{fig1}a is the lattice model of the normal-hexagonal semiconductor junction,
which consists of a square lattice at the left side and a hexagonal lattice at the right side.
The Hamiltonian of the whole system can be written as
\begin{align}\label{eq:a}
H=H_{L}+H_{R}+H_{T},
\end{align}
where $H_{L}$, $H_{R}$, $H_{T}$ are the Hamiltonians of the normal, hexagonal semiconductors,
and the coupling between them, respectively.
For the normal conductor, we consider the square lattice model with the dispersion relation of its carrier being quadratic. In the tight-binding representation, the Hamiltonian $H_{L}$ is of the form\cite{d1,d2}
\begin{align}\label{eq:b}
H_{L}=\sum_{i\sigma}\epsilon_{L} a^{\dagger}_{i\sigma} a_{i\sigma}+ \sum_{\langle{ij}\rangle \sigma} t_{L} a^{\dagger}_{i\sigma}a_{j\sigma},
\end{align}
where $a_{i\sigma}$ and $a^{\dagger}_{i\sigma}$ are the annihilation and creation operators
at the discrete site $i$, and $\epsilon_{L}$ is the on-site energy.
The second term in equation (\ref{eq:b}) is the nearest-neighbor hopping. $t_{L}$ is
the hopping energy, which is positive for valence bands and negative for conduction bands.
Experimentally, it is possible to break the A-B lattice symmetry in a hexagonal lattice like graphene.
For example, isolated graphene/BN bilayers break the chemical equivalence of graphene A and B lattice sites.\cite{E1,E2} Graphene growth on the reconstructed surface of MgO(111) also leads to A-B lattice symmetry breaking.\cite{E3,E4}
In addition, spin-orbit interaction can also play a very important role in some two-dimensional materials.
The emergence of the quantum spin Hall effect and topological insulators can be attributed to
the rise of the spin-orbit interaction.\cite{F1,F2,F3,F4}
Thus, we consider a general Hamiltonian at the hexagonal lattice side as
\begin{align}\label{eq:c}
H_{R}=\sum_{i\sigma}(\epsilon_{R}+\lambda_{i}\beta) b^{\dagger}_{i\sigma} b_{i\sigma}+ \sum_{\langle{ij}\rangle\sigma} t_{R} b^{\dagger}_{i\sigma}b_{j\sigma}\nonumber \\
+\sum_{\langle\langle{ij}\rangle\rangle \sigma \sigma^{'}}i\beta_{s}\nu_{ij}s_{\sigma \sigma^{'}}^{z}b_{i\sigma}^{\dagger}b_{j\sigma^{'}},
\end{align}
where $b_{i\sigma}$ and $b^{\dagger}_{i\sigma}$ are the annihilation and creation operators
at the discrete site $i$ of the right side. $\epsilon_{R}$ is the on-site energy,
and $\beta$ represents the energy difference between A-B sublattice.
Here $\lambda_{i}=\pm1$ for A(B)-sublattice.
The second term is the nearest-neighbor hopping term, and $t_{R}$ is the hopping energy.
The third term is the spin-orbit interaction which connects second nearest neighbor.
The same term also appears in the seminal work of the quantum spin Hall effect in graphene.\cite{F1}
$s_{z}$ is a Pauli matrix representing the electron's spin and $\nu_{ij}=-\nu_{ji}=\pm1$ depending on the orientation of the two site $i$ to $j$.\cite{F1}
The spin-orbit coupling $\beta_{s}$ is usually very small comparing to $t_{R}$,
but it can be in the order of meV by the Bi-cluster deposition.\cite{ref3}
The Hamiltonian $H_{T}$ of the coupling between the left and right lead is
\begin{align}\label{eq:d}
H_{T}=\sum_{ij\sigma}t_{c}a_{i\sigma}^{\dagger}b_{j\sigma}+h.c.
\end{align}
where $t_c$ is the coupling strength. Here we assume that the interface of the normal
and hexagonal semiconductors are perfect (see Fig.1a).
In usual, there are defects in the real devices and the interface is rough as well.
While the incident and refractive electrons are in the bottom of the conduction band
or the top of the valence band, their wavelengths are usually very long.
As long as the wavelength is much longer than the defect size and the interface roughness,
the direction of the refraction can be well maintained and the results are barely affected
by the defect and the imperfect interface.

The Brillouin zone of the hexagonal lattice is demonstrated in Fig.\ref{fig1}b.
We choose each unit cell containing four atoms (see red
box in Fig.\ref{fig1}a) which can simplify the calculations of the transmission coefficients,
hence the Brillouin zone is one half smaller than the usual Brillouin zone of graphene.
In Fig.\ref{fig1}c, we plot the energy bands along the dashed line in Fig.\ref{fig1}b with $k_y=0$.
Because of the Brillouin zone being half smaller, the number of the energy bands is double, and each spin orientation has four energy bands.
Breaking the A-B lattice symmetry induces a big energy gap at the K and K' points.
We mainly consider the electrons transport near the Fermi surface,
and the four lower energy bands can be ignored, as they are much below the Fermi surface $E=0$.
In Fig.\ref{fig1}d, the zoom-in figure of the energy bands near the K and K' points are presented.
The spin-orbit coupling can be viewed as an effective magnetic field $\bf{B}$ that points in the opposite directions at the K and K' points.
At K point, spin up electrons have lower energy while at K' point the other way around.
Thus, for energy bands with a definite spin orientation, the energy dispersion relation between the K and K' points are no longer equivalent any more.
In fact, this energy band is the same as that of TMDs in the normal phase.\cite{IsSCth}
So a junction consisting of a normal conductor coupled with the TMDs can be regarded
as a real example of our model. Double refraction occurs when electrons incident from normal conductor with the square lattice side
transmit to the two valleys, as demonstrated in the next section.

\begin{figure}[!htb]
\includegraphics[width=1.0\columnwidth]{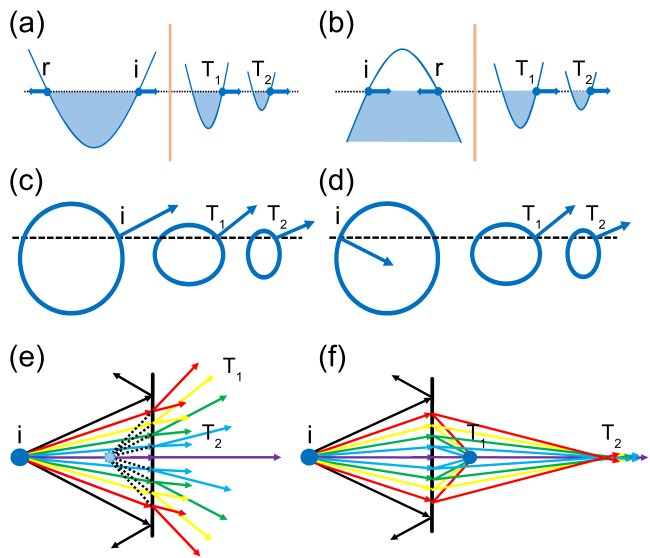}
\caption{(Color online) Heterojunction band diagrams and schematic of double refraction.
(\textbf{a}) The band diagrams of the n-n junction.
Incident electron i transmits to the two valleys with the
transmission coefficients $T_{1}$ and $T_{2}$.
(\textbf{b}) The corresponding p-n junction case.
(\textbf{c}) and (\textbf{d}) are the corresponding equienergy lines in the $k_x$-$k_y$ plane for
the n-n junction in (\textbf{a}) and the p-n junction in (\textbf{b}).
Here the arrows indicate the directions of the incident and refraction carriers.
(\textbf{e}) Schematic of the double positive refraction in the n-n junction. Electrons emitted
from the source i are bent different amounts at the n-n junction interface, forming two virtual focuses
at the same side of the electron source. Only one virtual focus is plotted in (\textbf{e}).
(\textbf{f}) Schematic of the double negative refraction in the p-n junction.
Electrons emitted from the source i undergo double negative refraction at the p-n junction interface,
forming two real focus at the opposite side of the electron source.}
\label{fig2}
\end{figure}

\section{\label{sec:Double refraction}Double refraction and double focusing}
In this section, we demonstrate how the normal-hexagonal semiconductor junctions leads
to the double refraction and double focusing.
Due to the Pauli matrix $\sigma_z$ commuting with the Hamiltonian in Eq.(\ref{eq:a}),
the spin in the $z$ direction is conserved in the scattering process.
So in the following analysis, we first consider a spin component, e.g. spin-up one.
Fig.\ref{fig2}a and \ref{fig2}b respectively illustrate the band diagrams of
the n-n junction and p-n junction for a spin component,
and the corresponding equienergy lines in the $k_x$-$k_y$ plane are shown in Fig.\ref{fig2}c and \ref{fig2}d.
For an incident carrier (electron or hole) from the left normal conductor, there are two
beams of outgoing electrons, due to the two non-equivalent valleys.
Notice that the directions of the two beams of outgoing electrons are usually different.
So the double refraction occurs in this normal-hexagonal semiconductor junction.
Because of the transverse translation invariance in the model,
the transverse component of the momentum ($k_y$) is conserved when electrons transmit across the junction.
So the transverse group velocity preserves the sign between the n-n junction (see Fig.\ref{fig2}c),
but changes a sign between the valence bands in the p side and the conduction bands in the n side (see Fig.\ref{fig2}d).
As a consequence, the double refraction is positive in a n-n junction,
while negative in a p-n junction.
Fig.\ref{fig2}e and \ref{fig2}f are the schematic diagrams of the positive and
negative double reflections, respectively.
For the positive double refraction, the outgoing electrons
are bent different amounts as they pass through the junction and form two virtual focuses
at the same side of the electron source i.
However, in Fig.\ref{fig2}f, two real focuses are formed at the opposite side of the electron
source, attributing to the opposite transverse velocity and the negative refractive index.

\begin{figure}[!htb]
\includegraphics[width=1.0\columnwidth]{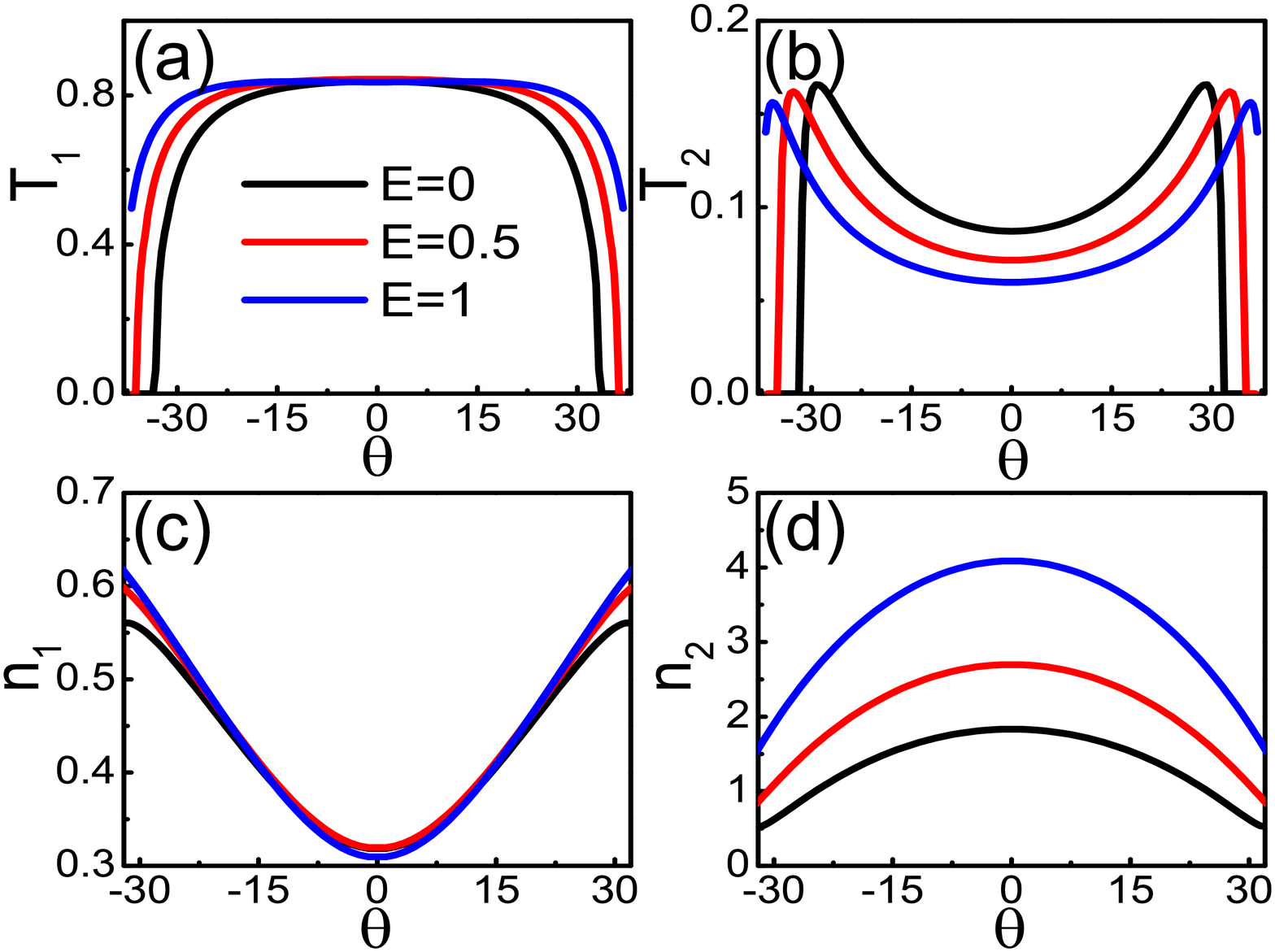}
\caption{(Color online) Transmission coefficients and refractive indices for the n-n junction.
(\textbf{a}) and (\textbf{b}) are the transmission coefficients $T_{1}$ and $T_{2}$
at the two valleys as a function of the incident angle $\theta$ for the different incident energy $E$.
(\textbf{c}) and (\textbf{d}) are the corresponding refractive indices.
The parameters are $\epsilon_{L}=7$, $t_{L} =-8$, others are the same as that in Fig.\ref{fig1} .}
\label{fig3}
\end{figure}

To further investigate the nature of the double refraction, we calculate the transmission coefficients and
the refractive indices for both case in detail.
By applying the non-equilibrium Green's function method to the tight-binding Hamiltonian $H$
in equation (\ref{eq:a}), the transmission coefficients $T_1$ and $T_2$
can be obtained quite straightforward by following the same procedure as in Refs.[\onlinecite{G2,G1}].
Fig.\ref{fig3} displays the case of the double positive refraction in a n-n junction,
where (a) and (b) are the transmission coefficients and (c) and (d) are the refractive indices.
For small incident angle, the transmission coefficient $T_{1}$ is nearly a constant,
and is insensitive to the incident energy $E$.
As the incident angle increases, $T_{1}$ rapidly reduces and $T_{2}$ significantly increases.
For large incident angle, both $T_{1}$ and $T_{2}$ drop to zero, attributing to the absence of the
corresponding transmition modes at the n side. Thus, electrons undergo total reflection at large incident angle.
Note that $T_{1}$ dominates the whole scattering process for a large range of parameters.
The increasing of the incident energy enhence $T_{1}$ at large incident angle,
but meanwhile suppress the transmission coefficient $T_{2}$.
The total transmission probability $T=T_{1}+T_{2}\approx 1$ indicates that the n-n junction is
almost transparent for the electrons.
Meanwhile, the two transmission modes have different refractive property as shown in Fig.\ref{fig3}c and \ref{fig3}d.
Here the refractive index is define as $n_{i}=\sin{\theta}/\sin{\theta_{i}}$, where $\theta$ is the incident angle
and $\theta_{i}$ ($i=1,2$) is the refraction angle of transmission mode $i$.
The refractive index $n_{1}$ is insensitive to the incident energy, and increases as the incident angle grows.
We emphasize that $n<1$ for a large range of parameters,
hence the electronic refraction at this valley serves as optically thinner medium in optics.
For a perfect virtual focus, the refractive index n obeys the law $n^2=c+(1-c)\sin^2\theta$,
where the parameter $0<c<1$ in optically thinner medium.
The refractive index $n_{1}$ approximately follows this law, hence this virtual focus is nearly perfect
(see Fig.\ref{fig2}e).
For the refraction at the other valley, the refractive index $n>1$ for a large range of parameters,
hence electronic refraction at this valley serves as optically denser medium in optics.
Here $c>1$ in the optically denser medium. Although $n_{2}$ is sensitive to incident energy,
it also approximate matches the law $n^2=c+(1-c)\sin^2\theta$.
Therefore, the second virtual focus is also relatively good
(The second virtual focus is not plotted in Fig.\ref{fig2}e).

\begin{figure}[!htb]
\includegraphics[width=1.0\columnwidth]{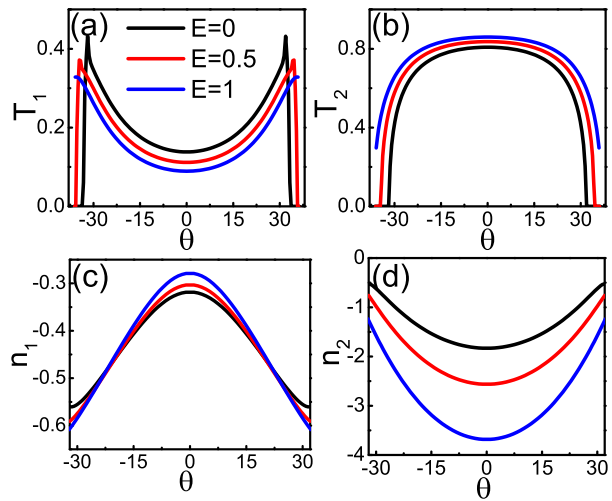}
\caption{ (Color online) Transmission coefficients and refractive indices for the p-n junction.
(\textbf{a}) and (\textbf{b}) are the transmission coefficients $T_{1}$ and $T_{2}$ at the two valleys
versus the incident angle $\theta$ for the different incident energy $E$.
(\textbf{c}) and (\textbf{d}) are the corresponding refractive indices.
The parameters are $\epsilon_{L}=-7$, $t_{L} =8$, others are the same as that in Fig.\ref{fig1} .}
\label{fig4}
\end{figure}

Next, we focus on the p-n junction in which the double negative refraction occurs,
as illustrated in Fig.\ref{fig4}.
The main features of the transmission coefficients are basically the same as the case of positive refraction, except the position of $T_{1}$ and $T_{2}$ are changed.
In this case, $T_{2}$ dominates the scattering process (see Fig.\ref{fig4}a and \ref{fig4}b),
but $T_{1}$ is not very small and it has the same order of $T_2$.
As for the refractive indices shown in Fig.\ref{fig4}c and \ref{fig4}d,
$n_{1}$ and $n_{2}$ become negative and almost have a mirror symmetry with the positive refractive
indices in Fig.\ref{fig3}c and \ref{fig3}d about the axis $n=0$.
Contrary to the previous case, two real focus are formed at the opposite side of the electron source,
and the focusing performance of the incident electrons is pretty good (see Fig. \ref{fig2}f).

The spin-down electrons show similar refractive behaviors at the two valleys. The double refraction is positive
for a n-n junction and negative for a p-n junction, and the absolute value of the refractive index $|n|>1$ at one valley and $|n|<1$ at the other.
Because of the time reversal invariance of our Hamiltonian, the scattering matrix of the spin-up and spin-down electrons
satisfy the relation ${\bf s}_{\uparrow}^{T}={\bf s}_{\downarrow}$, which indicates that $T_{1\uparrow}+T_{2\uparrow}=T_{1\downarrow}+T_{2\downarrow}$.

\begin{figure}[!htb]
\includegraphics[width=1.0\columnwidth]{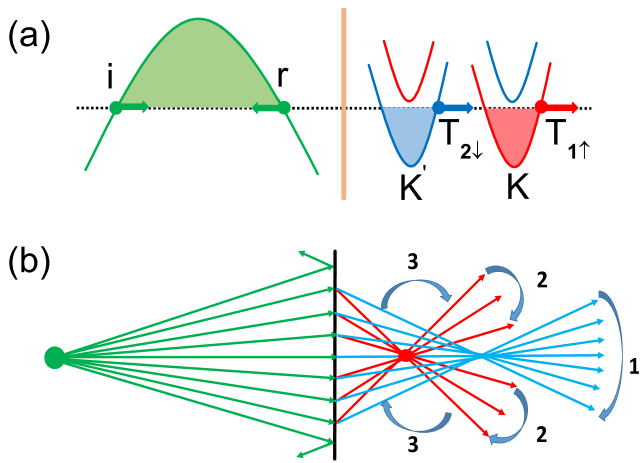}
\caption{(Color online) A spin splitter based on the normal-hexagonal p-n junction.
(\textbf{a}) is the band diagrams of the spin splitter based on the normal-hexagonal p-n junction.
Here the Fermi energy $E$ locates between the energy bands of the up-spin and down-spin species
of the n side.
(\textbf{b}) is the schematic diagram of the electron's trajectory in the spin splitter.
Electrons with different spin are separated in real space and focus in different regions. }
\label{fig5}
\end{figure}

\section{\label{sec:Spin splitter}Spin splitter and detective device}

Spin-orbit coupling modifies the electronic band structures at the two valleys and
the double refraction become spin dependent as mentioned above.
Thus, for spin non-polarized incident electrons from the left side,
the outgoing electrons on the right side will be spin-polarized,
i.e. the normal-hexagonal semiconductor junction can be worked as a spin splitter.
Below, as an example, we consider the p-n junction system, and the energy bands are shown in Fig.\ref{fig5}a.
Here the red lines denote the energy bands of spin-up electrons on the n side while the blue lines denote spin-down ones, and the green lines are the bands on the p side where the spin-up and spin-down electrons are degenerate.
In order to clearly illustrate the physical picture of the spin splitter,
the Fermi energy $E$ is set to locate between the red line and blue line (see Fig.\ref{fig5}a).
In this case, the incident spin-up electrons can only get through the junction via valley K at the state $T_{1\uparrow}$
while the spin-down electrons via valley K' at the state $T_{2\downarrow}$.
Notice the states $T_{1\uparrow}$ and $T_{2\downarrow}$ are not mutually the time-reversal states and
the dispersion relations at the two valleys $\epsilon_{K\uparrow}({\bf k}-{\bf K}) \not= \epsilon_{K'\downarrow}({\bf k}-{\bf K'})$,
although the system has the time-reversal invariance with
the dispersion relations having $\epsilon_{K\uparrow}({\bf k}-{\bf K}) = \epsilon_{K'\downarrow}(-{\bf k}-{\bf K'})$.
So the refractive indices of the two valleys are different, leading that the electrons with
different spin are separated in real space and focus in different regions.
By detailed calculation of the incident and refraction angles,
we plot the schematic diagram of the electron's trajectory in Fig.\ref{fig5}b.
Spin-up electrons focus in a small area while spin-down electrons focus in a relatively large area.
The electron's outgoing area can be divided in three regions.
In region 1, both spin-up and spin-down electrons can be observed,
but spin-down electrons are in the majority.
In region 2, only spin-up outgoing electrons could be observed.
In region 3, we can hardly detect any outgoing electron.
In particular, at two focal points the density of the spin-up and spin-down outgoing electrons are very high,
which can be detected by using ferromagnetic STM.
Note that here the spin-up and spin-down electrons are at the valleys K and K', respectively. So
the spin splitter is also a valley splitter .

\begin{figure}[!htb]
\includegraphics[width=1.0\columnwidth]{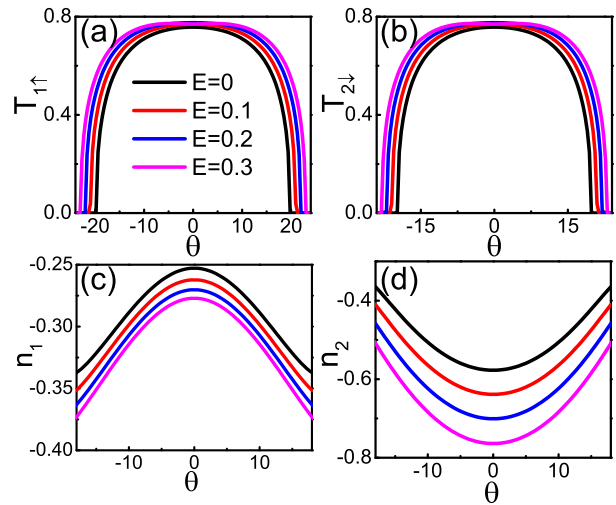}
\caption{(Color online) Transmission coefficients and refractive indices for the spin splitter.
(\textbf{a}) and (\textbf{b}) are the transmission coefficients $T_{1\uparrow}$ and $T_{2\downarrow}$ for spin-up and spin-down electrons, respectively.
(\textbf{c}) and (\textbf{d}) are the corresponding refractive indices.
The parameters are $\epsilon_{R}=-4.2$, $t_{R} =8$, $\beta=4$, $\beta_{s}=-0.1$, $\epsilon_{L}=-7$ and $t_{L} =8$.}
\label{fig6}
\end{figure}

Let us study the transmission coefficients and refractive indices of the spin splitter.
Fig.\ref{fig6}a and \ref{fig6}b are the transmission coefficients $T_{1\uparrow}$ and $T_{2\downarrow}$ of
the two transport channels.
The two transmission coefficients are in fact identical for various incident energy.
This is becasue for each spin there is only one transmission mode, and the general relation
$T_{1\uparrow}+T_{2\uparrow}=T_{1\downarrow}+T_{2\downarrow}$ reduces to $T_{1\uparrow}=T_{2\downarrow}$
in present case.
$T_{1\uparrow}$ and $T_{2\downarrow}$ are the maximum when the incident angle $\theta=0$.
With the increase of $\theta$, they decrease. And at large $\theta$,
$T_{1\uparrow}$ and $T_{2\downarrow}$ drops rapidly to zero.
However, due to the band structures at the two valleys being not completely the same,
the refractive indices present different behavior as illustrated in Fig.\ref{fig6}c and \ref{fig6}d.
With larger incident angle, the refractive index $n_{1}$ decreases, while $n_{2}$ shows opposite behavior.
The refractive index $n_{1}$ approximately follows the asymptotic behaviour of perfect lenses,
and the spin-up electrons focus in a small area.
However, $n_{2}$ obviously do not follow the law, and the spin-down electrons focus in a relatively large area.
Note that by changing the spin-orbit coupling constant $\beta_{s}$,
different types of focusing pattern can be realised.

\section{\label{sec:conclusions}Summary}
In conclusion, we have proposed a model to realize double refraction
and double focusing of electric current using a normal-hexagonal semiconductor junction.
By breaking the A-B lattice symmetry and introducing the Rashba spin-orbit interaction,
a large energy gap is formed and spin-up and spin-down electrons experience opposite effective magnetic field at the two valleys.
Incident electrons transmit to the two valleys, and result in a double refraction at
the hexagonal lattice side.
The double refraction can be either positive or negative, depending on the junction being n-n type or p-n type.
The two valleys reveal different type of refractive behaviors.
The absolute value of the refractive index $|n|>1$ at one valley and $|n|<1$ at the other.
Additionally, the p-n junction can be used as a spin splitter (valley splitter),
which could be verified by the ferromagnetic STM.
Our results may be useful for the engineering of double
lenses in electronic system and have underlying application in spintronics.

\section*{Acknowledgement}
This work was financially supported by National Key R and D Program of China (2017YFA0303301),
NBRP of China (2015CB921102) and NSF-China under Grants Nos. 11574007 and 11274364.

\end{document}